\documentclass{article}
\usepackage{spconf,amsmath,graphicx,hyperref}
\usepackage{booktabs}
\usepackage{vntex}      
\usepackage[utf8]{inputenc}
\usepackage{booktabs}
\usepackage{subcaption}
\usepackage{caption}

\title{How Far Do SSL Speech Models Listen for Tone?\\Temporal Focus of Tone Representation under Low-resource Transfer} 
\name{Minu Kim, Ji Sub Um, Hoirin Kim}
\address{School of Electrical Engineering, KAIST, Daejeon, Republic of Korea}

\begin{document}
%
\maketitle
\begin{abstract}
Lexical tone is central to many languages but remains underexplored in self-supervised learning (SSL) speech models, especially beyond Mandarin. We study four languages with complex and diverse tone systems (Burmese, Thai, Lao, and Vietnamese) to ask how far such models “listen” for tone and how transfer operates in low-resource conditions. As a baseline reference, we estimate the temporal span of tone cues: approximately 100\,ms (Burmese/Thai) and 180\,ms (Lao/Vietnamese). Probes and gradient analysis on fine-tuned SSL models reveal that tone transfer varies by downstream task: automatic speech recognition fine-tuning aligns spans with language-specific tone cues, while prosody- and voice-related tasks bias toward overly long spans. These findings indicate that tone transfer is shaped by downstream task, highlighting task effects on temporal focus in tone modeling.
\end{abstract}


\section{Introduction}
\label{sec:intro}

Lexical tone is a defining feature of many of the world’s languages, accounting for about 40\%~\cite{wals}, or up to 60–70\% with pitch-accent systems~\cite{yip2002tone}.
As a core linguistic cue for distinguishing word meanings, tone is also vital in speech technology: failing to model it degrades recognition and synthesis, while accurate encoding improves robustness and intelligibility~\cite{liang2025tone,coto2021explicit,jiang2024semantic,tao2024toneunit}. 
Tone is thus central not only in linguistic typology but also in the design of robust speech systems.

Meanwhile, recent advances in self-supervised learning (SSL) for speech have improved how speech is modeled and represented~\cite{mohamed2022self}. 
These models have been shown to encode rich segmental information, particularly phonemes~\cite{ji2022predicting}, and have achieved strong results in multilingual automatic speech recognition (ASR) and related tasks~\cite{lodagala2024all,zhuang2024svq}. 
However, supra-segmental features such as lexical tone have received far less attention. 
Existing analyses have largely focused on Mandarin or a few other languages~\cite{shen2024encoding,de2024layer,gogoi2025tone}, while systematic study of Southeast Asian languages is still lacking, despite their complex tone systems with dynamic pitch movements~\cite{brunelle2016tone}. Since their tone cues unfold over time, this highlights the importance of analyzing temporal spans in tone modeling.

We thus focus on four Southeast Asian languages, relying on temporal tone cues and encompassing diverse families: Burmese, Thai, Lao, and Vietnamese, which remain under-studied and relatively low-resource in speech technology. We first estimate the optimal temporal span for tone classification with a log-Mel spectrogram and logistic regression, finding shorter spans for Burmese and Thai ($\approx$100 ms) and longer ones for Lao and Vietnamese ($\approx$180 ms). We then probe SSL models layer by layer to analyze tone-centered gradient activity, and evaluate how different fine-tuning strategies (e.g., ASR in the same or different languages, prosody- or voice-related tasks such as emotion/gender classification, and vanilla pretrained models) transfer to tone recognition under low-resource settings by aligning with these spans. 

Our contributions are threefold. 
(1) We provide a systematic analysis of the temporal span needed for tone recognition in tonal languages. 
(2) We present a gradient-based layer-wise probe to examine how SSL models capture tone spans. 
(3) We show that fine-tuning on tonal ASR yields representations most consistent with these spans. 
Taken together, our results demonstrate that tone representations are transferable in SSL models when downstream finetuning aligns with tone spans.

\section{Models and Data}
\label{sec:models-data}

\subsection{Models}

\textbf{Base architectures.}
We analyze a set of representative self-supervised speech encoders. 
Our primary focus is on \textit{wav2vec~2.0 large}~\cite{baevski2020wav2vec} and its multilingual extension \textit{XLS-R 300M}~\cite{babu2021xls}. 
For additional comparison, we also include \textit{MMS 300M}~\cite{pratap2024scaling} and \textit{mHuBERT-147}~\cite{zanon2024mhubert}. 

\vspace{5pt}

\noindent\textbf{Fine-tuning settings.} To examine how downstream adaptation shapes tone representations, we consider 
single-language fine-tuned variants of \textit{XLS-R 300M}. 
For ASR, models are separately fine-tuned on Burmese, Thai, Lao, and Vietnamese (FLEURS~\cite{conneau2023fleurs}), as well as on English and Mandarin (CommonVoice v22.0~\cite{ardila2019common}). 
For prosody- and voice-related tasks, we use models fine-tuned on emotion recognition (RAVDESS~\cite{livingstone2018ryerson}), gender classification (LibriSpeech train-clean-100~\cite{panayotov2015librispeech}), and speaker verification (VoxCeleb1~\cite{nagrani2017voxceleb}).
For comparison, we include both vanilla and target-language ASR fine-tuned \textit{MMS 300M} and \textit{mHuBERT-147}.

\subsection{Languages and Data}

\begin{figure}[t]
    \centering
    \includegraphics[width=\linewidth]{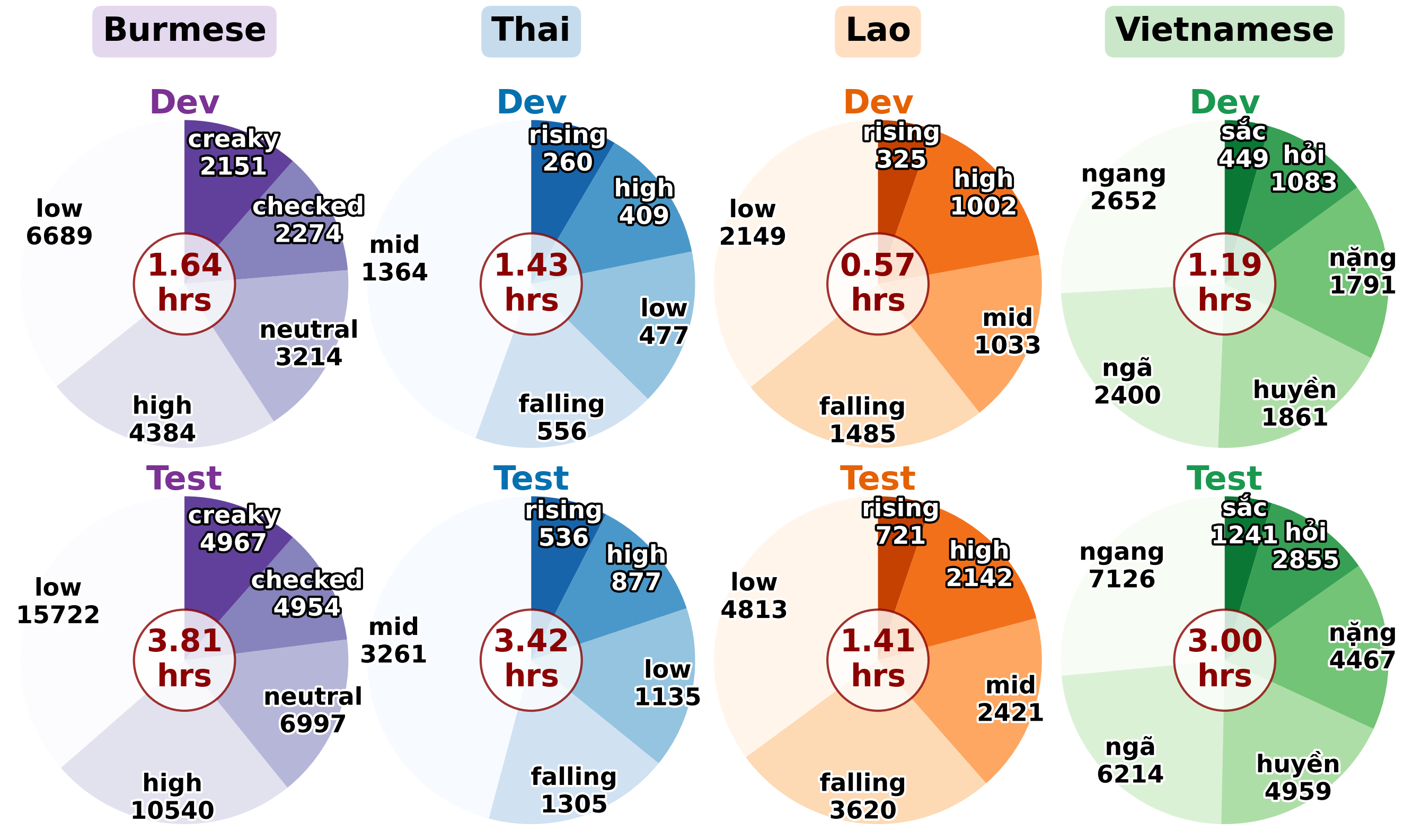}
    \caption{Tone distributions across languages. 
    (Vietnamese tones: ngang (high
level), huyền (low level), sắc (rising), hỏi (falling-rising), ngã (broken), nặng (falling)~\cite{pham2003key}.)}
    \label{fig:lang-stats}
\end{figure}

\textbf{Languages.} 
We analyze four tonal languages using the FLEURS corpus \cite{conneau2023fleurs}: Burmese (Sino-Tibetan), Thai (Tai-Kadai), Lao (Tai-Kadai), and Vietnamese (Austroasiatic). 
To ensure no data overlap, we fine-tune ASR models on the \textit{train} splits, while the tone-related probing classifiers are trained on the \textit{dev} sets and evaluated on the \textit{test} sets.
Figure~\ref{fig:lang-stats} summarizes the tone distributions and dataset sizes across languages, highlighting their data scarcity and the need for representation transfer in low-resource settings~\cite{kim2025improving}.

\vspace{5pt}

\noindent\textbf{Tone alignment.}
To obtain time-aligned tone units, we first apply language-specific grapheme-to-phoneme (G2P) conversion 
(Burmese: burmese-G2P\footnote{\url{https://github.com/kyaw-yethu/burmese-G2P}}; 
Thai, Lao, Vietnamese: espeak-based phonemizer~\cite{bernard2021phonemizer}) and extract tone labels. 
We then perform CTC-based forced alignment~\cite{graves2006connectionist} with wav2vec 2.0 models fine-tuned on the same corpus, 
in line with established token-level alignment procedures~\cite{kurzinger2020ctc,zhu2022phone}.

\section{Analysis Methods}
\label{sec:methods}

\subsection{Span Analysis with Acoustic Features}

Before analyzing self-supervised models, we establish a baseline of the temporal resolution required for tone recognition.  
Lexical tones can rely on different temporal cues, ranging from brief pitch movements to longer F$_0$ trajectories~\cite{yang2017duration,perkins2024interaction}, which motivates estimating the span at which tone distinctions become reliably detectable across languages.  
To this end, we train logistic regression classifiers on log-Mel features using fixed windows (20–300\,ms).  
Performance across window sizes, evaluated with macro-F1, yields the optimal span per language, providing a reference point for interpreting how SSL model representations capture tone.

\subsection{Probing SSL Representations}

To examine how tone information is localized in SSL models, we adopt a gradient-based sensitivity analysis~\cite{simonyan2013deep,li2016understanding}. 
We first train simple linear probes on each frozen SSL layer, evaluating their performance with macro-F1: for each tone segment, the hidden state at the center frame is used to predict its tone label. 
We then compute input gradients with respect to the gold-class logit $z^{(\ell)}_{y}$ at layer~$\ell$, yielding squared gradient energies aligned to input time frames (Eq.~\ref{eq:grad}).

\begin{equation}
g^{(\ell)} = \frac{\partial z^{(\ell)}_{y}}{\partial x}, 
\qquad
E^{(\ell)}(t) = \bigl(g^{(\ell)}_t\bigr)^2 .
\label{eq:grad}
\end{equation}

Energies are aligned to time offsets $\Delta t = |t - t_c|/f_s$ relative to the tone center $t_c$, where $f_s$ is the sampling rate (16\,kHz), and binned (20\,ms up to 1000\,ms) to form normalized histograms $H^{(\ell)}(\Delta t)$. 
From these, we derive the center-of-mass radius $r_{\mathrm{com}}^{(\ell)}$ (in ms; Eq.~\ref{eq:rcom}), summarizing the average temporal focus of gradient sensitivity. 
Smaller $r_{\mathrm{com}}^{(\ell)}$ indicates sharper temporal concentration around tone centers.

\begin{equation}
r_{\mathrm{com}}^{(\ell)} = 
\frac{\sum_{\Delta t} \Delta t \, H^{(\ell)}(\Delta t)}
     {\sum_{\Delta t} H^{(\ell)}(\Delta t)} .
\label{eq:rcom}
\end{equation}

\begin{figure}[t]
    \centering
    \includegraphics[width=\linewidth]{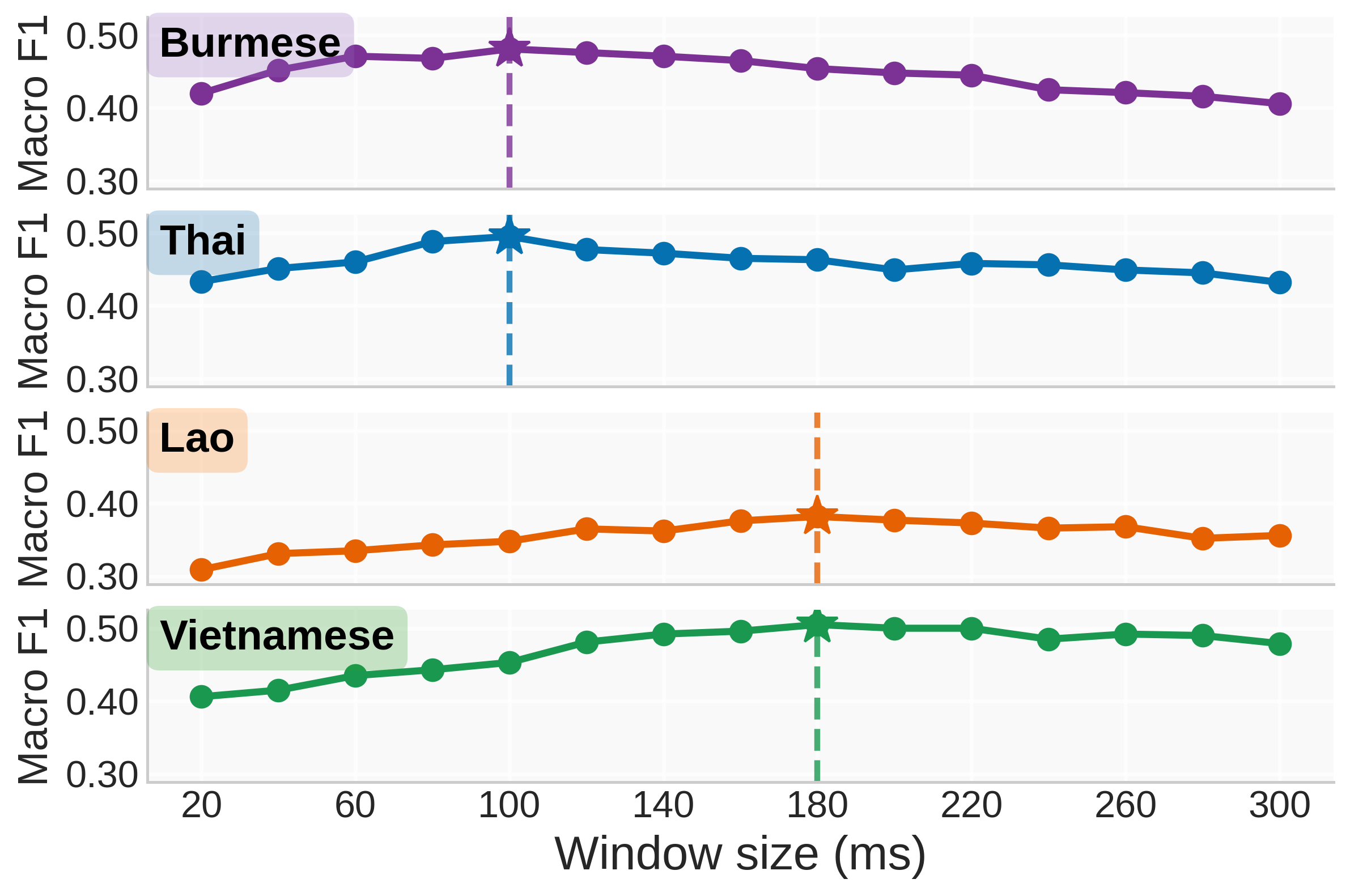}
    \caption{Baseline tone classification macro-F1 with logistic regression across varying window lengths (20--300\,ms). 
    Shorter spans suffice for Burmese and Thai, whereas Lao and Vietnamese require longer spans.}
    \label{fig:span-baseline}
\end{figure}

\begin{figure*}[t]
  \centering
  \includegraphics[width=\textwidth]{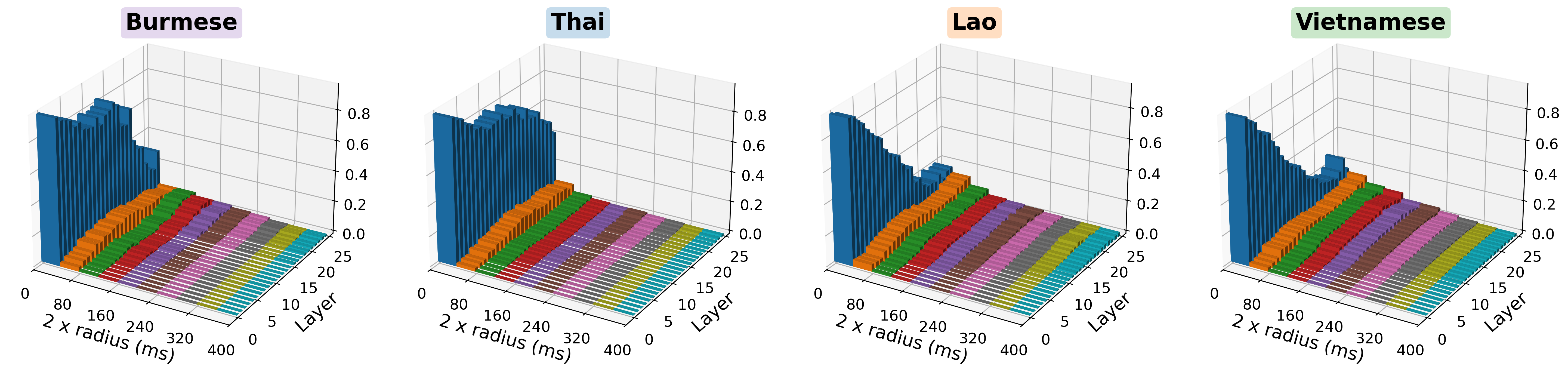}
\caption{Normalized gradient energy around tone centers with XLS-R ASR models fine-tuned in each language. 
Burmese/Thai show sharper focus, while Lao/Vietnamese show broader spreads, consistent with acoustic span baselines.}

  \label{fig:grad-sensitivity}
\end{figure*}

\begin{figure*}[t]
    \centering
    \begin{subfigure}[t]{0.48\linewidth}
        \centering
        \includegraphics[width=\linewidth]{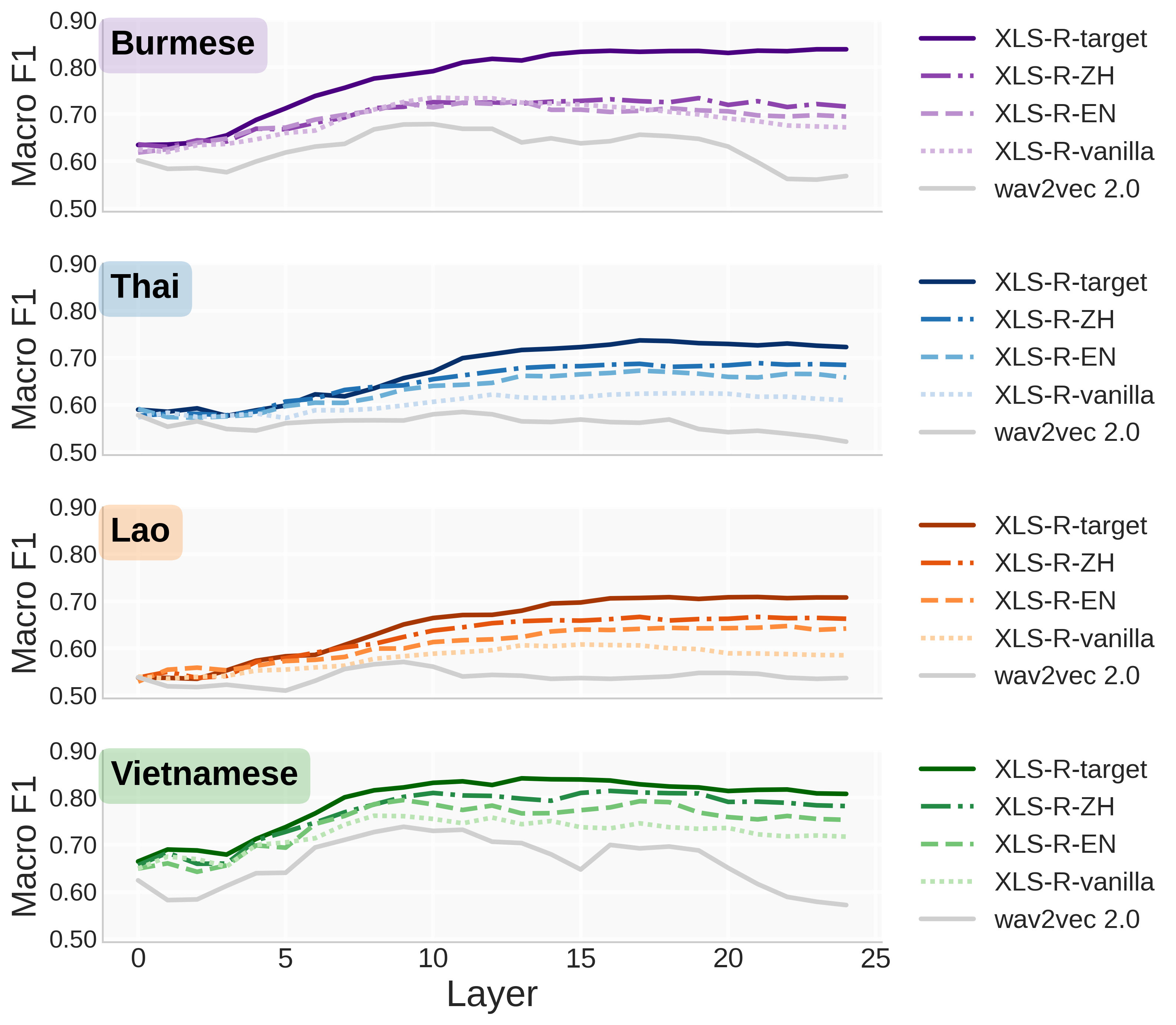}
    \end{subfigure}
    \hfill
    \begin{subfigure}[t]{0.48\linewidth}
        \centering
        \includegraphics[width=\linewidth]{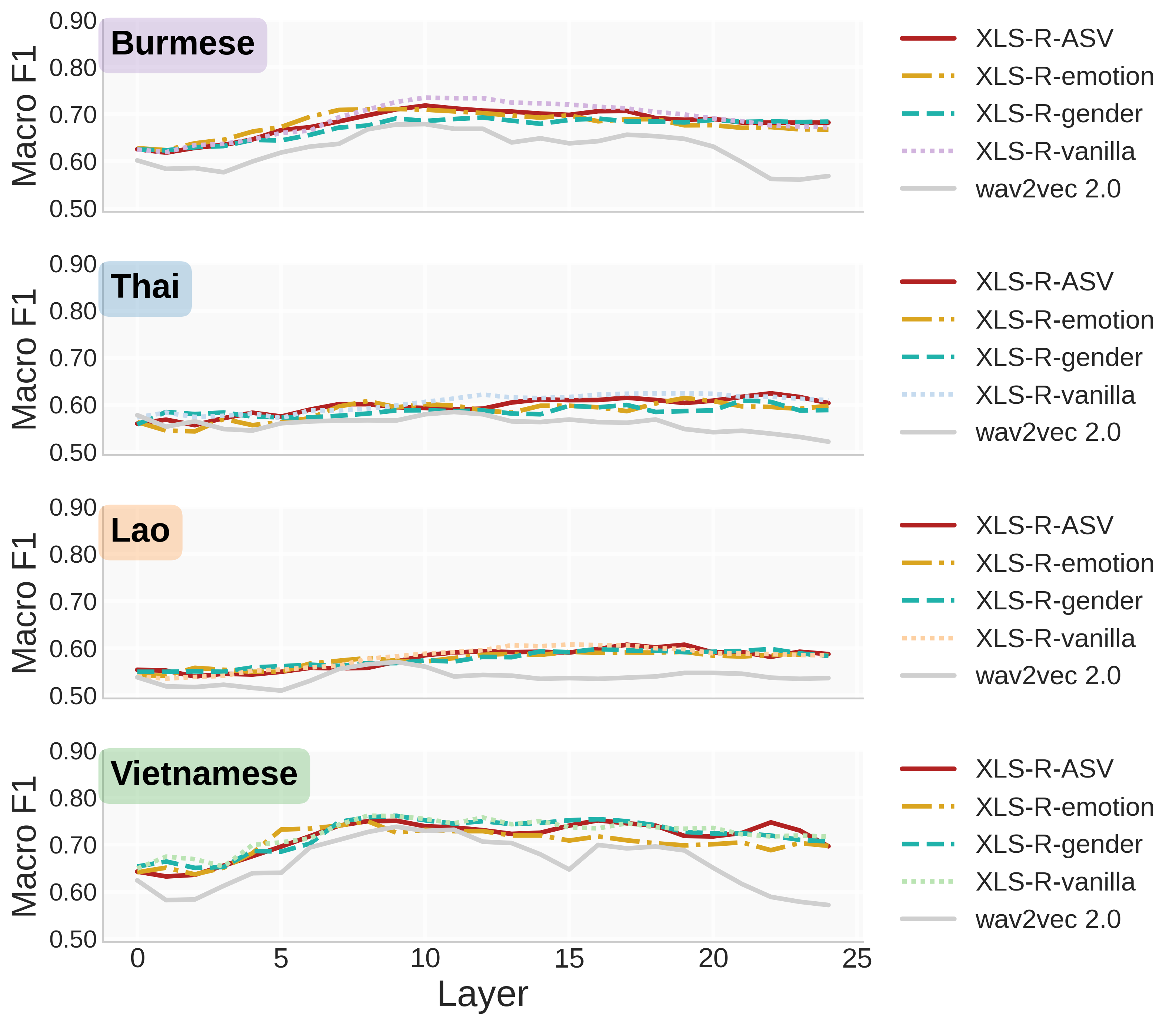}
    \end{subfigure}
    \caption{Layer-wise probe performance (macro-F1) across Burmese, Thai, Lao, and Vietnamese using different SSL models (XLS-R-target: ASR on target language; XLS-R-ZH: ASR on Mandarin Chinese; XLS-R-EN: ASR on English; XLS-R-ASV: speaker verification; XLS-R-emotion: emotion recognition; XLS-R-gender: gender classification). Probe performance usually peaks in mid-to-high layers, showing that lexical tone information is mainly captured at those layers.}
    \label{fig:probe-accuracy-combined}
\end{figure*}

\section{Results}
\label{sec:results}

\subsection{Baseline Span Analysis}
\label{subsec:span}
We begin by quantifying the temporal resolution required for tone recognition using acoustic features. 
Figure~\ref{fig:span-baseline} shows tone classification macro-F1 as a function of window length. 
Burmese and Thai perform best with spans around 100\,ms, while Lao and Vietnamese require longer spans of roughly 180\,ms. 
Performance declines when windows grow beyond these ranges, confirming that the effective tone span is language-specific and providing a reference scale for evaluating SSL model representations.


\subsection{Gradient-based Temporal Sensitivity}
\label{subsec:sens}

Figure~\ref{fig:grad-sensitivity} shows normalized gradient energy around tone centers for Burmese, Thai, Lao, and Vietnamese, 
using XLS-R models fine-tuned for ASR in each language. 
Gradients are measured within a 1000\,ms window and normalized across time; for readability, only the inner 200\,ms radius ($\approx$400\,ms total span) is displayed. 
The x-axis marks temporal offset from the tone center, the y-axis encoder layers, and the z-axis the proportion of gradient energy.

Across languages, clear differences emerge: Burmese and Thai show sharp concentration around tone centers, indicating narrow focus, 
whereas Lao and Vietnamese display a bit broader spreads, especially in higher layers, suggesting longer-span tone cues.
These differences mirror the acoustic baseline: languages with inherently short-span tone cues produce tighter sensitivity, while those with longer cues induce wider temporal receptive fields. 
Overall, the findings suggest that ASR fine-tuning adapts SSL representations to match the language-specific temporal scope of tone cues.

\begin{figure*}[t]
  \centering
  \includegraphics[width=\textwidth]{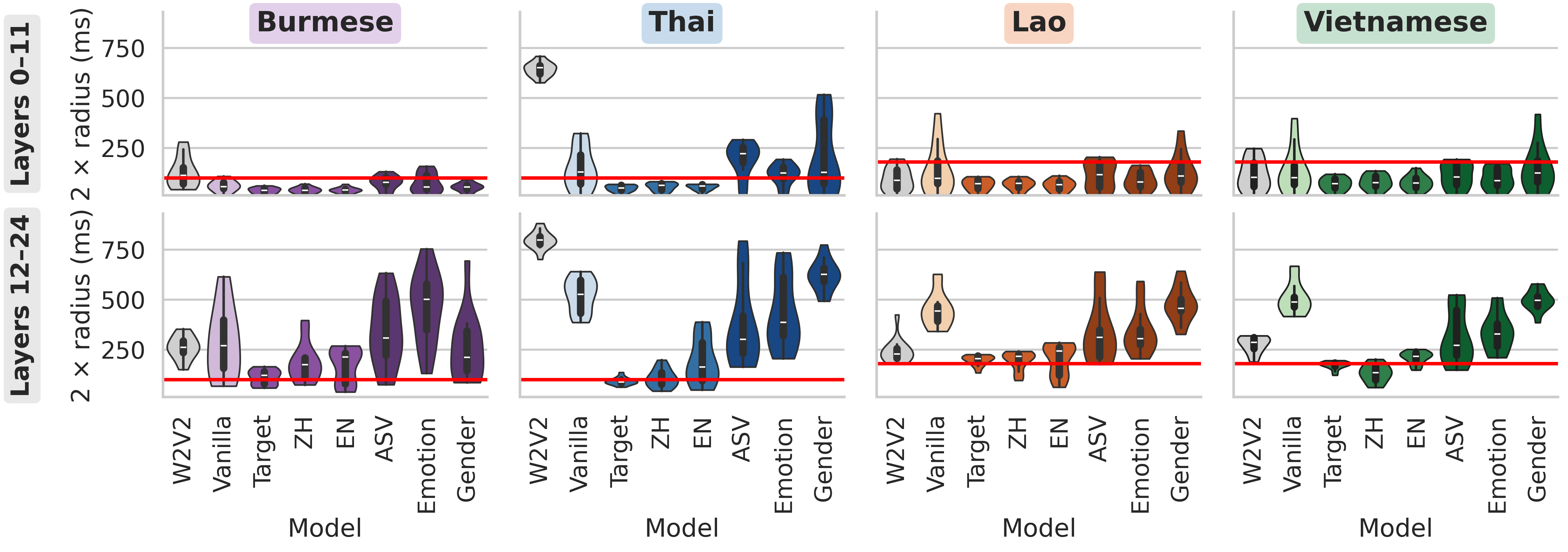}
\caption{Distributions of layerwise effective span ($2\times r_{\mathrm{com}}^{(\ell)}$) for tone gradients, shown separately for lower (0–11) and higher (12–24) layers across models (W2V2: wav2vec 2.0 large; Vanilla: XLS-R-vanilla; ZH: XLS-R-ZH; EN: XLS-R-EN; ASV: XLS-R-ASV; Emotion: XLS-R-emotion; Gender: XLS-R-gender). Red line indicates baseline span; target ASR shows the closest fit.}

  \label{fig:radius}
\end{figure*}

\begin{figure}[t]
    \centering
    \includegraphics[width=\linewidth]{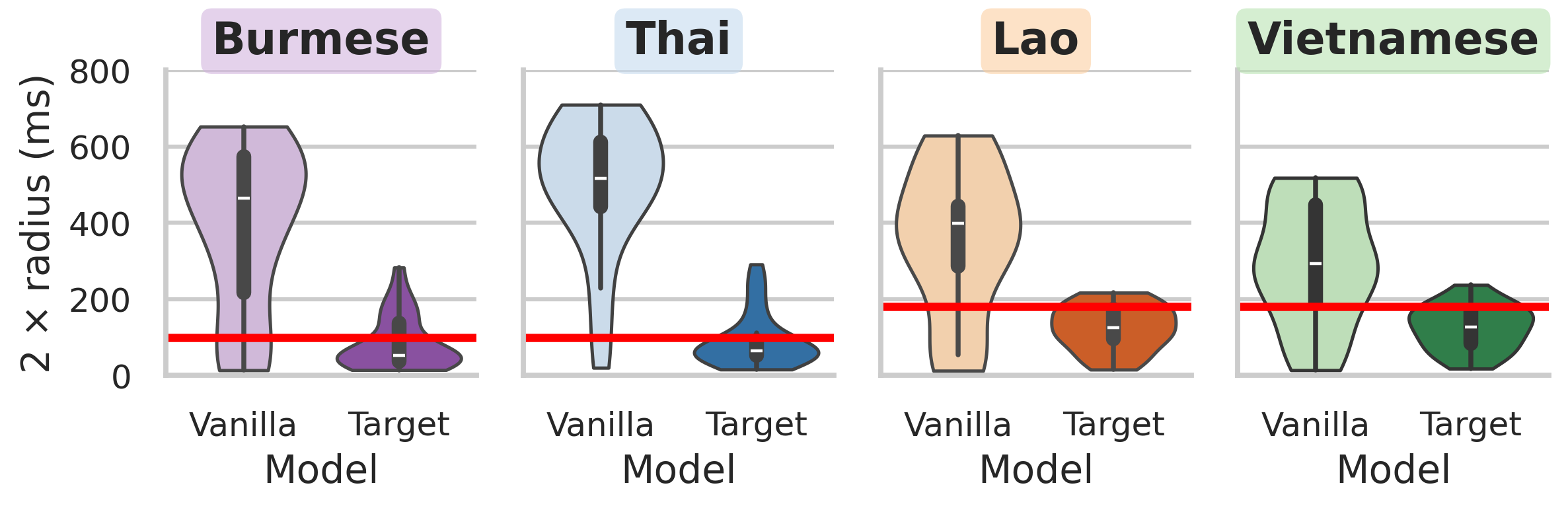}
    \caption{Distributions of tone gradient spans along layers in MMS models. Target ASR fine-tuning yields span alignment.}
    \label{fig:mms-radius}
\end{figure}

\subsection{Effect of Fine-tuning Task}

\textbf{Layer-wise probe performance.} 
Figure~\ref{fig:probe-accuracy-combined} compares the effect of different fine-tuning strategies on probe performance. 
ASR fine-tuning consistently yields the highest layer-wise F1 scores, with the advantage becoming most pronounced in the middle to higher encoder layers where tone cues are captured most effectively. 
Among the cross-lingual settings, Mandarin ASR fine-tuning, which itself involves lexical tone, achieves the second-best results, while English ASR fine-tuning still improves macro-F1 over the XLS-R vanilla model. 
In contrast, fine-tuning for prosody- or voice-related tasks (emotion, gender, speaker) produces little to no improvement: their probe performance is nearly indistinguishable from the vanilla model, indicating that these objectives do not effectively separate tone information.

\vspace{5pt}

\noindent\textbf{Layer-wise gradient sensitivity.} 
Building on the analysis in subsection~\ref{subsec:sens}, we quantify layerwise gradient concentration using the effective span ($2\times r_{\mathrm{com}}^{(\ell)}$) around tone centers.
Figure~\ref{fig:radius} shows results for XLS-R under different fine-tuning settings.
We separate lower (0–11) and higher (12–24) layers since tone information is more salient in higher layers.
Target-language ASR fine-tuning yields the most stable localization: in higher layers, spans match the language-specific baselines (red lines) from subsection~\ref{subsec:span} (100\,ms; 180\,ms), 
whereas early layers lack such concentration, explaining weaker tone probe results in 0-11. 
Cross-lingual ASR shows similar trends, with Mandarin aligning more closely than English.
In contrast, prosody- and voice-related fine-tuning pushes spans overly long, reflecting their broad acoustic focus and explaining lower probe scores.
Overall, span alignment is a key factor for capturing tone information effectively.

\begin{figure}[t]
    \centering
    \includegraphics[width=\linewidth]{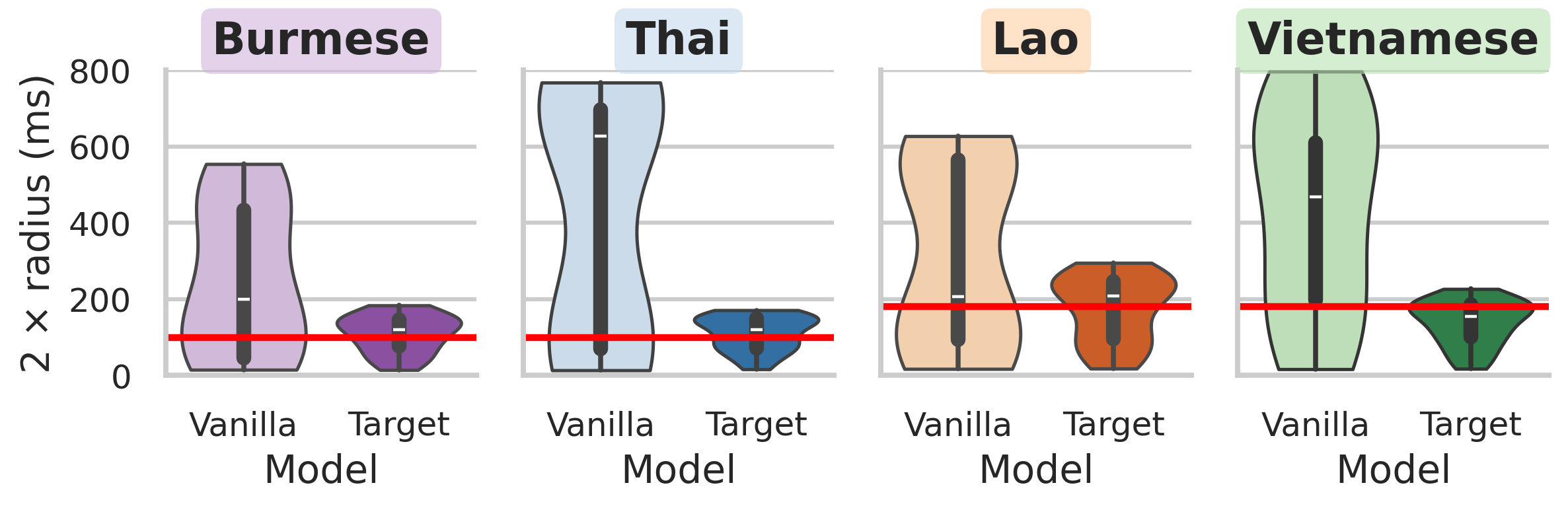}
    \caption{Distributions of tone gradient spans in mHuBERT models. Target ASR fine-tuning enhances tone localization.}
    \label{fig:mhubert-radius}
\end{figure}

\subsection{Comparison with Other Models}

Figures~\ref{fig:mms-radius} and~\ref{fig:mhubert-radius} show results for MMS~\cite{pratap2024scaling} and mHuBERT-147~\cite{zanon2024mhubert}. 
For simplicity, we report radii across the full depth of each model.
Both models broadly mirror the patterns observed with XLS-R: Vanilla models localize tone weakly, while target-language ASR fine-tuning sharpens span alignment to language-specific cues.

\section{Conclusion}
\label{sec:conclusion}

We analyzed tone representations in SSL models for four Southeast Asian languages, showing that mid-to-high layers capture tone most effectively and that transfer is strongest when fine-tuning tasks align with the language-specific span of tone cues. 
These findings highlight tone as a transferable suprasegmental feature, providing a guide for seed model selection and performance tuning in low-resource settings.

\vfill\pagebreak
\footnotesize
\section{Acknowlegements}
\vspace{-7pt}
This work was supported by Institute of Information \& communications Technology Planning \& Evaluation (IITP) grant funded by the Korea government(MSIT) (No.RS-2025-02215393).

\bibliographystyle{IEEEbib}
\bibliography{refs}

\end{document}